\documentclass[a4paper,12pt]{article}

\usepackage[cp1250]{inputenc} 
\usepackage{authblk}
\usepackage{graphicx}
\usepackage{amsmath}
\usepackage{amsthm}
\usepackage{a4wide}
\begin{document}

\title{Can \textit{Google Trends} search queries contribute to risk diversification?\footnote{Published as Kristoufek, L. (2013): Can \textit{Google Trends} search queries contribute to risk
diversification? \textit{Scientific Reports} \textbf{3}, 2713; DOI:10.1038/srep02713. The paper is available at http://www.nature.com/srep/2013/130919/srep02713/full/srep02713.html.}}

\author[a,b]{Ladislav Kristoufek}
\affil[a]{Institute of Economic Studies, Faculty of Social Sciences, Charles University in Prague, Opletalova 26, 110 00, Prague, Czech Republic, EU}
\affil[b]{Institute of Information Theory and Automation, Academy of Sciences of the Czech Republic, Pod Vodarenskou Vezi 4, 182 08, Prague, Czech Republic, EU, tel.: +420266052243}
\date{\today}
\maketitle

\begin{abstract}
\footnotesize
Portfolio diversification and active risk management are essential parts of financial analysis which became even more crucial (and questioned) during and after the years of the Global Financial Crisis. We propose a novel approach to portfolio diversification using the information of searched items on \textit{Google Trends}. The diversification is based on an idea that popularity of a stock measured by search queries is correlated with the stock riskiness. We penalize the popular stocks by assigning them lower portfolio weights and we bring forward the less popular, or peripheral, stocks to decrease the total riskiness of the portfolio. Our results indicate that such strategy dominates both the benchmark index and the uniformly weighted portfolio both in-sample and out-of-sample.
\end{abstract}

\normalsize

In the most recent years, social sciences have obtained access to huge datasets based on internet activity of millions of users all over the world. This way, it has become possible to analyze complex behavior of internet users describing various patterns, regularities and connections to other (real world) phenomena. Among the most frequently utilized providers of data, social media such as Twitter and Facebook \cite{Miller2011,Metaxas2012} and searching engines Google and Yahoo! \cite{Vosen2011,Choi2012,Bordino2012} play the most important role. This has brought the social sciences closer to the natural sciences, methods of which have repeatedly proved worthy providing a whole new perspective on analyzed phenomena \cite{Buldyrev2010,Fehr2002,Gabaix2003,Haldane2011,Kenett2012a,Kenett2012,King2011,Lazer2009,Lillo2003,Lux1999,Petersen2010,Pozzi2013,Pozzi2008,Preis2012,Preis2011,Schweitzer2009,Sornette2011,Vespignani2009}. In finance and economics, \textit{Google Trends} have drawn a special attention as the frequency of searched terms has been shown to provide useful information for nowcasting and forecasting of various phenomena ranging from trading volumes \cite{Preis2010} through consumer behavior \cite{Vosen2011,Goel2010,Carri`ere-Swallow2013} to macroeconomic variables \cite{Fondeur2013,Preis2012a} and finance \cite{Mondaria2010,Drake2012,Preis2013}.

Specifically in finance, it has been shown that internet search queries can help partially explain home bias in investment allocation \cite{Mondaria2010}, search volume for a company name is positively correlated with transaction volume of the corresponding stock on both weekly \cite{Preis2010} and daily time scales \cite{Bordino2012}, the information diffusion around the earnings announcements of publicly traded companies is shown not to be instantaneous with the announcement release \cite{Drake2012}, trading strategies based on search frequency of finance-related terms produce profits above the ones of random strategies \cite{Preis2013} and \textit{Wikipedia} activity is argued to be linked to the upcoming financial critical events \cite{Moat2013}. In financial theory, financial analysis and financial management, portfolio diversification and risk optimization play an essential role \cite{Markowitz1952,Sharpe1964,Lintner1965,Merton1972} so that profit maximization is practically pushed aside due to its theoretical unfeasibility in the efficient markets hypothesis framework \cite{Samuelson1965,Fama1965,Fama1970}. 

Here, we contribute to the literature by expanding the utility of information provided by \textit{Google Trends} search queries to the portfolio diversification. By partially building on the previous results in the field indicating search volumes to be correlated and to be able to help in predictions of volume and variance of financial assets and also on classical economic literature discussing link between traded volume and variance \cite{Karpoff1987}, we propose a portfolio diversification strategy based on the search volume of stock-related terms. The diversification strategy stems in an idea that the more frequently the stock-related term is searched for the higher the risk (in the financial perspective) of the specific stock. Therefore, such a popular stock should be discriminated with a low weight in the final portfolio to decrease the total risk of the portfolio. Reversely, the least popular stocks are given a higher weight in the portfolio.

\section*{Results}

To analyze the performance of the search volume based portfolio selection strategy, we utilize the search queries provided by \textit{Google Trends}. We analyze two types of queries. First, we focus on the effect of searching the ticker symbol of a stock. Second, we use the combination of the word ``stock'' and the ticker symbol to ensure that the searched term is not misinterpreted as the ticker symbol. DJI (Dow Jones Industrial Average, or Dow Jones 30) index and its 30 components create the initial dataset. For each of the 30 stocks, we construct the series of weekly returns $r_{i,t}$ defined as
\begin{equation}
r_{i,t}=\frac{p_{i,t}-p_{i,t-1}}{p_{i,t-1}}
\end{equation}
for stock $i$ at week $t$ where $p_{i,t}$ is the closing price of stock $i$ at the end of week $t$.

Diversification strategy is constructed as follows. To discriminate for the popularity of the stock, we propose a power-law rule to obtain portfolio weights. Let $V_{i,t}$ be a search volume of stock-related term of stock $i$ at week $t$. The weight $w_{i,t}$ of stock $i$ in the portfolio at time $t$ is defined as
\begin{equation}
\label{eq:weights}
w_{i,t}=\frac{V_{i,t}^{-\alpha}}{\sum_{j=1}^{N}{V_{j,t}^{-\alpha}}}
\end{equation}
where $N$ is the number of stocks in the portfolio and $\alpha$ is a power-law parameter measuring the strength of discrimination of a stock being too frequently searched for. The normalization factor $\sum_{j=1}^{N}{V_{j,t}^{-\alpha}}$ ensures that $\sum_{i=1}^{N}{w_{i,t}}=1$ for all $t$. Note that if $\alpha>0$, the more frequently looked up stocks are assigned a lower weight, and if $\alpha<0$, the more searched for stocks are preferred in the portfolio. For $\alpha=0$, a uniformly diversified portfolio with $w_{i,t}\equiv w = \frac{1}{N}$ for all $i$ and $t$ is retrieved. We opt for the power-law discrimination rule to ensure that even highly popular stocks are still at least marginally present in the portfolio. Therefore, we do discriminate the popular stocks but we do not want to have their weights in the portfolio vanish too quickly and too frequently, which would be the case with e.g. an exponential discrimination rule.

We are interested in performance of the proposed methodology both in- and out-of-sample. The former is a standard approach to measure quality of portfolio optimization but the latter is more useful for practical applicability of the diversification strategy. The in-sample performance is then based on portfolio weights rebalancing at week $t$ according to Eq. \ref{eq:weights} and gaining/losing in the same period. The out-of-sample is based on rebalancing at the end of week $t$ and realizing gains/losses at the end of week $t+1$. Structure and availability of search query data enables such rebalancing (this issue is discussed in more detail in the Methods section). We are mainly interested in two portfolio performance measures -- standard deviation and the Sharpe ratio (defined as the ratio between average return and standard deviation). Standard deviation is a standard measure of risk and the Sharpe ratio represents the standardized average return of the portfolio. In the financial literature, the portfolio is standardly constructed to either minimize the risk or maximize the Sharpe ratio.

As noted earlier, we analyze two types of searched terms -- the ticker symbol (e.g. GE for General Electric Company or XOM for Exxon Mobil Corporation) and the combination of the word ``stock'' and ticker symbol (e.g. ``stock GE'' and ``stock XOM''). Even though the \textit{Google Trends} database ranges back to the beginning of 2004, we analyze two different periods for the two approaches -- 3.1.2005 - 24.6.2013 (443 weeks) for the former and 5.1.2009 - 24.6.2013 (234 weeks) for the latter -- due to data availability (zero-value observations are very frequent for the dates before the analyzed periods). We also had to omit tickers AA, BA, BAC, CAT, DD, DIS, HD, KO, MCD, PG, T, TRV, UNH, VZ for the first approach due to infrequent queries but also ambiguity and interchangeability of the ticker symbols with stock unrelated terms and abbreviations. For the second approach, we had to erase observations for AXP, MRK, TRV, UNH, UTX due to infrequent search queries.

\begin{figure}
\center
\begin{tabular}{cc}
\includegraphics[width=3in]{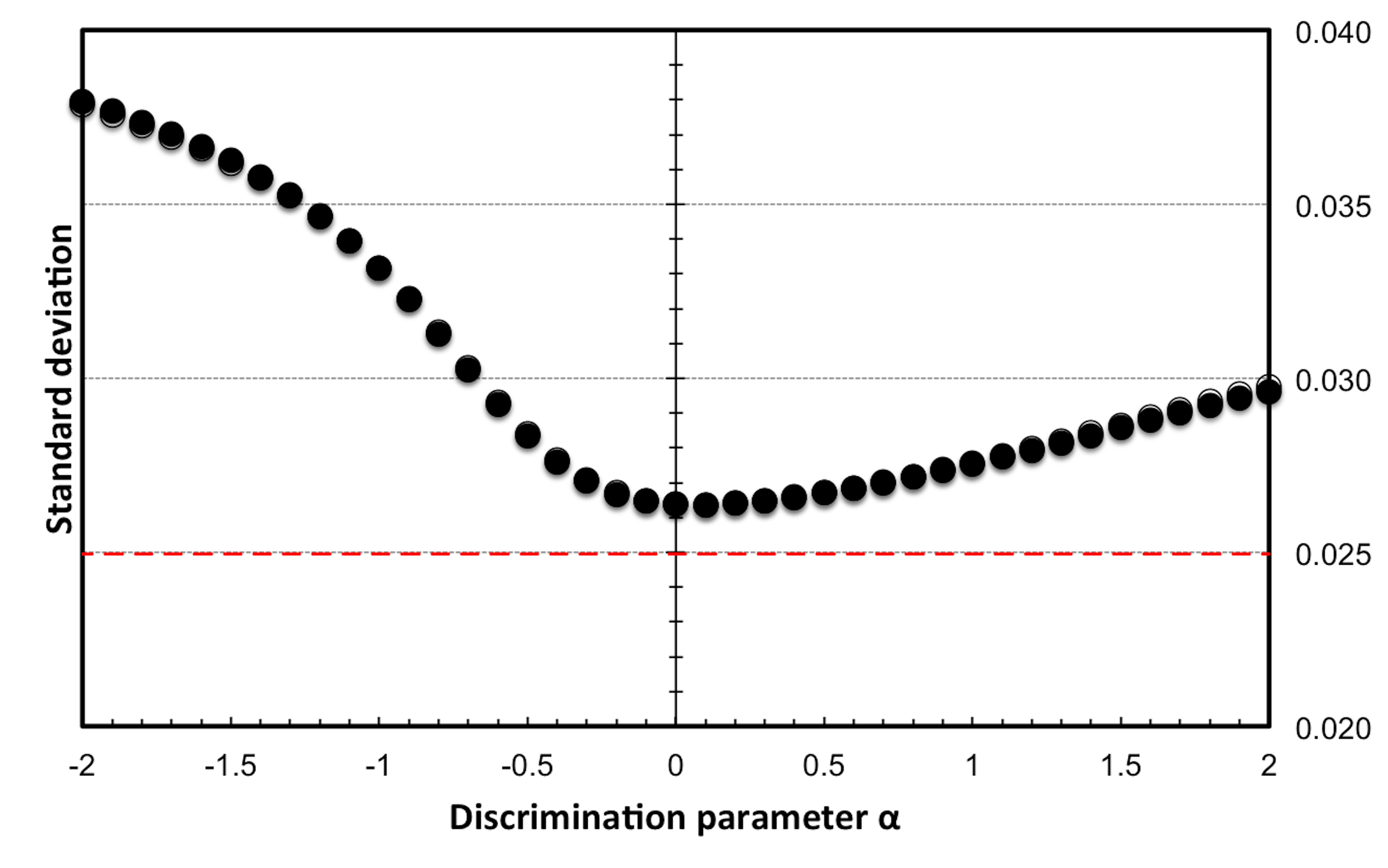}&\includegraphics[width=3in]{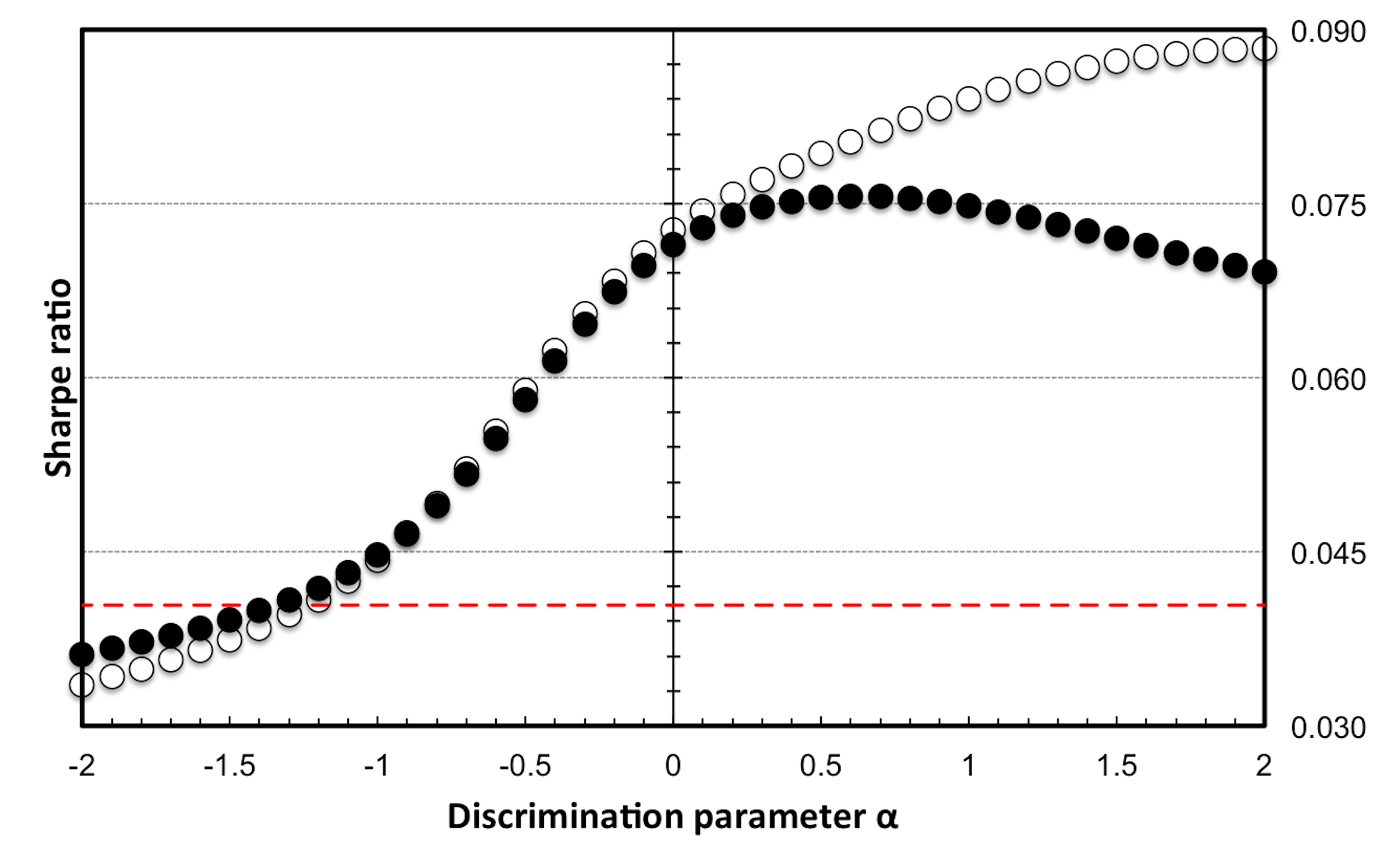}\\
\end{tabular}
\caption{\footnotesize\textbf{Portfolio performance based on ticker symbols.} Standard deviation (\textit{left}) and Sharpe ratio (\textit{right}) are shown for in-sample (full symbols) and out-of-sample (empty symbols) performances of constructed portfolios. The discrimination parameter $\alpha$ ranges between -2 and 2 with a step of 0.1. The middle point ($\alpha=0$) represents the uniformly weighted portfolio. The red dashed line represents the benchmark DJI index. Minimum variance portfolio is found for $\alpha=0.1$ for both in- and out-of-sample approaches. In general, the resulting standard deviations for varying $\alpha$ practically overlap for both approaches. For the Sharpe ratio measure, the performance differs. For the in-sample, the Sharpe ratio is maximized for $\alpha=0.6$, and for the out-of-sample, it is maximized for $\alpha=2$. \label{fig1}}
\end{figure}

In Fig. \ref{fig1}, we show standard deviations and Sharpe ratios for portfolios based on the ticker approach for $\alpha$ varying between -2 and 2 with a step of 0.1. For reference, we also show the performance of the DJI index. Both in-sample and out-of-sample performances are shown in the figure. The behavior of the standard deviation is practically identical for the in-sample and out-of-sample -- it decreases with increasing $\alpha$ between $\alpha=-2$ and $\alpha=0.1$ where the deviation reaches its minimum, and for $\alpha>0.1$, it increases again. To see whether the higher standard deviations are offset by increasing return, we report the behavior of Sharpe ratio as well.  For the in-sample, the ratio increases with $\alpha$ and reaches its maximum at $\alpha=0.6$. For the out-of-sample, the ratio increases steadily up to $\alpha=2$. Importantly, the maximum ratio of the out-of-sample (approximately 0.09) is higher than the one for the in-sample (approximately 0.075) by around 20\%, which is something rather unexpected. In standard case, we expect the out-of-sample performance to be inferior to the in-sample performance. 

These results yield several interesting implications. First, the \textit{Google Trends} based strategies are able to reach lower risk level than the uniformly weighted portfolio. Second, the standardized returns of these portfolios also outperform the uniformly distributed ones. Third, the strategies are successful even in the out-of-sample which is probably the most important result. Fourth, comparing the out-of-sample performance of the \textit{Google Trends} based strategy with the passive buy-and-hold strategy (buying the DJI index at the beginning of 2005, holding and selling it at the end of June 2013), the search-based strategy outperforms the DJI strategy strongly (0.09 Sharpe ratio of the query-based strategy compared to less than 0.045 of the DJI index). And fifth, the portfolio selections with positive $\alpha$ strongly outperform the ones with negative $\alpha$ implying that the discrimination based on the search popularity pays off in the portfolio selection. 

\begin{figure}
\center
\begin{tabular}{cc}
\includegraphics[width=3in]{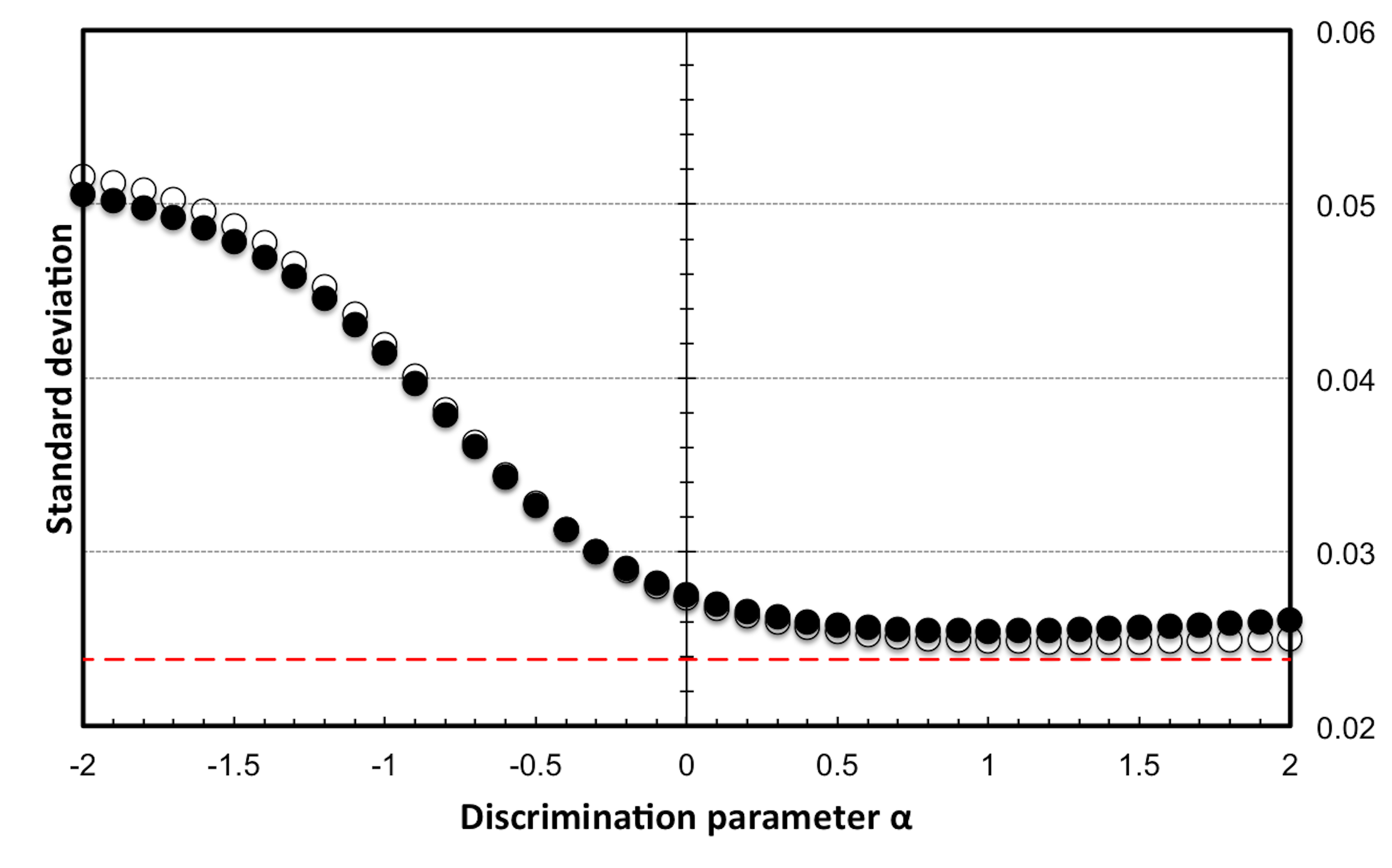}&\includegraphics[width=3in]{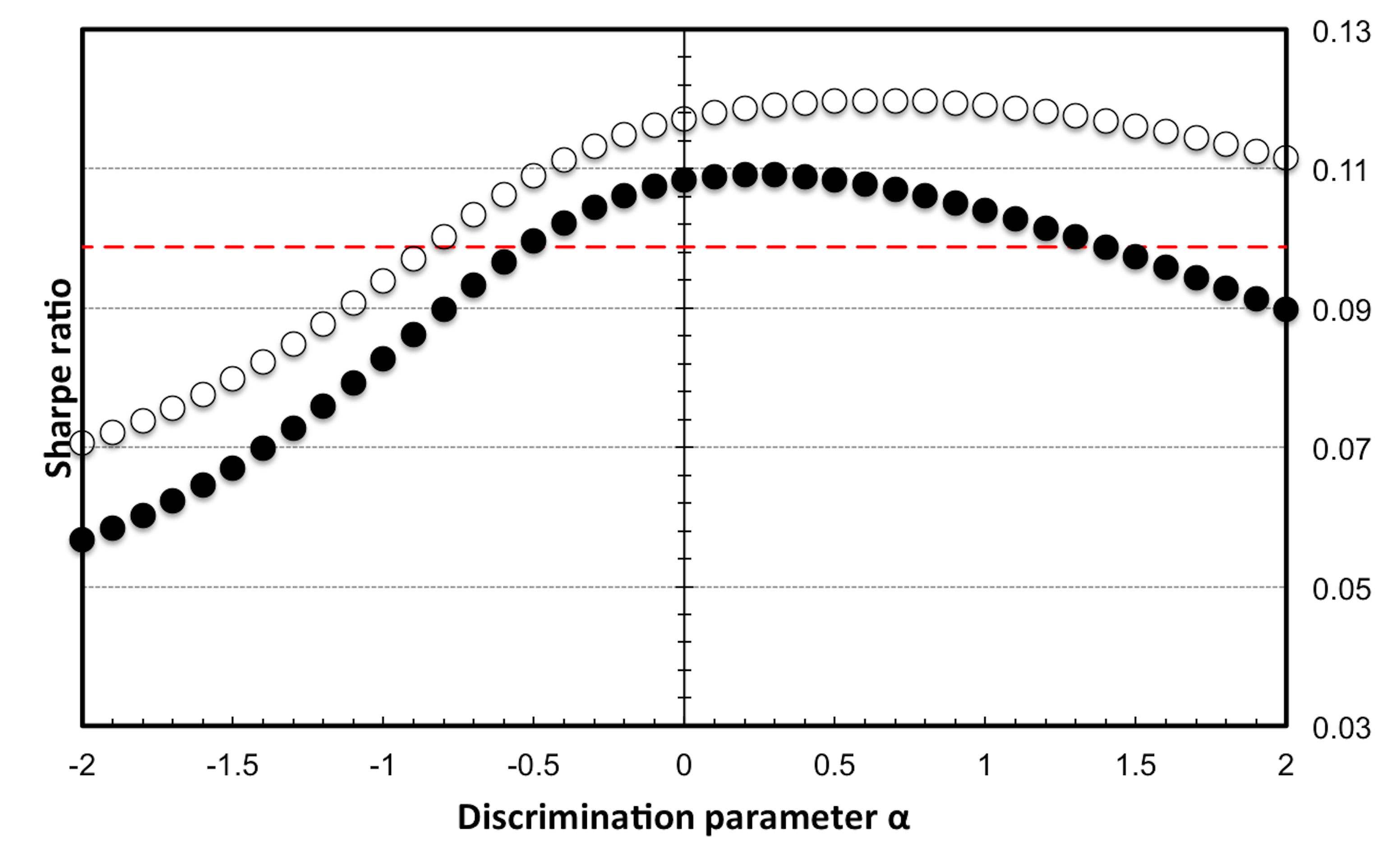}\\
\end{tabular}
\caption{\footnotesize\textbf{Portfolio performance based on ticker symbols combined with the word ``stock''.} Standard deviation (\textit{left}) and Sharpe ratio (\textit{right}) are shown for in-sample (full symbols) and out-of-sample (empty symbols) performances of constructed portfolios. The discrimination parameter $\alpha$ ranges between -2 and 2 with a step of 0.1. The middle point ($\alpha=0$) represents the uniformly weighted portfolio. The red dashed line represents the benchmark DJI index. Minimum variance portfolio is found for $\alpha=1$ and $\alpha=1.3$ for the in- and out-of-sample approaches, respectively. Again, the resulting standard deviations for varying $\alpha$ are very close for both approaches. For the Sharpe ratio measure, the performance again differs. The Sharpe ratio is maximized for $\alpha=0.2$ and $\alpha=0.6$ for the in- and out-of-sample, respectively. \label{fig2}}
\end{figure}

Qualitatively the same results are observed for the stock/ticker approach. In Fig. \ref{fig2}, we observe that standard deviation of the portfolio decreases with the discrimination factor between $\alpha=-2$ and $\alpha=1$ (in-sample), and between $\alpha=-2$ and $\alpha=1.3$ (out-of-sample), and then increases slowly. Again, there exists a search-based portfolio with lower variance than the uniformly weighted portfolio. For Sharpe ratio, we again observe that the out-of-sample dominates the in-sample performance, which only supports the utility of the search-based approach for investment strategies. The maximum ratios are located at $\alpha=0.2$ and $\alpha=0.6$ for the in-sample and the out-of-sample, respectively. The passive DJI strategy is again dominated but not as strongly as in the previous case.

\section*{Discussion}

\begin{figure}
\center
\begin{tabular}{cc}
\includegraphics[width=3in]{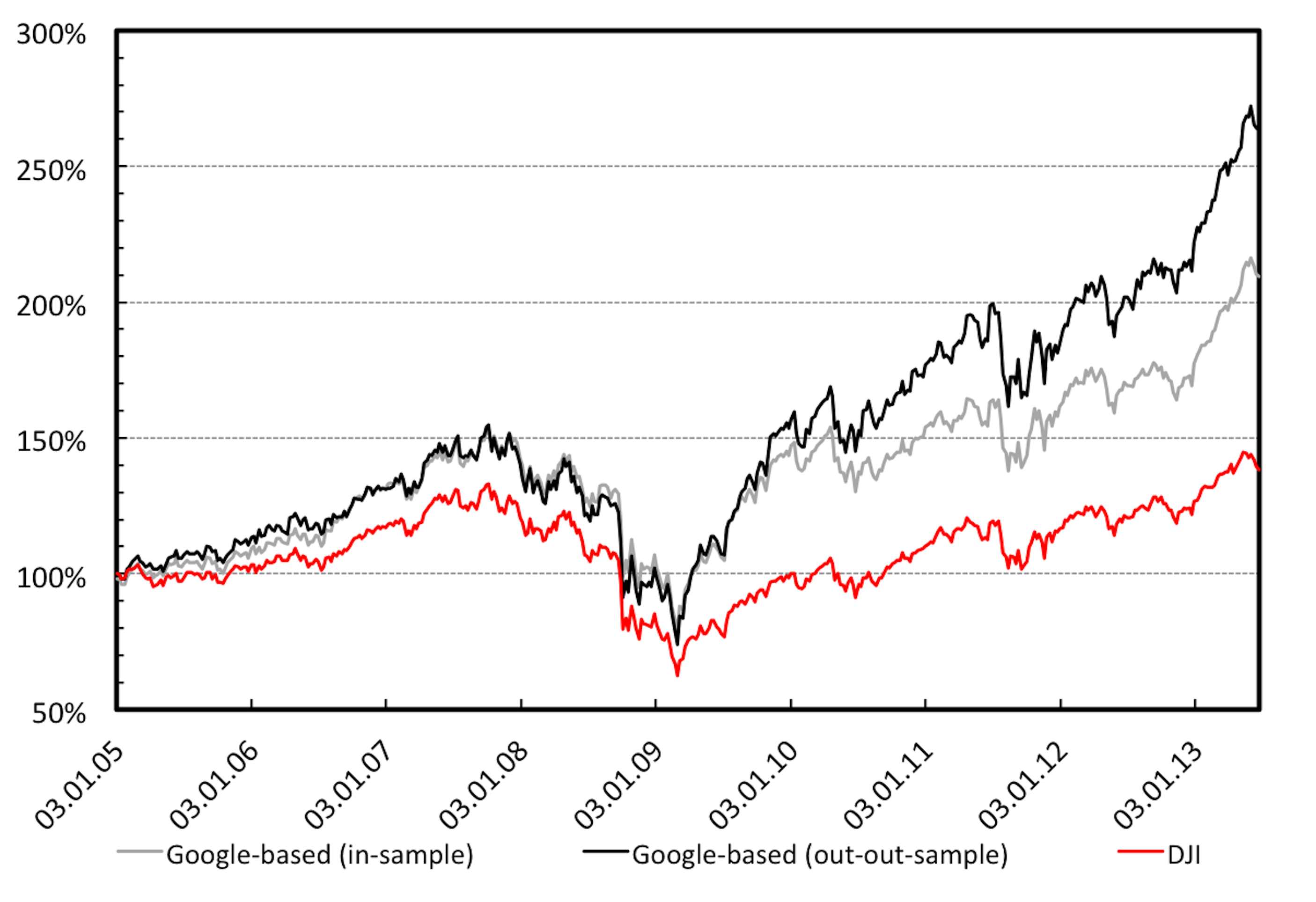}&\includegraphics[width=3in]{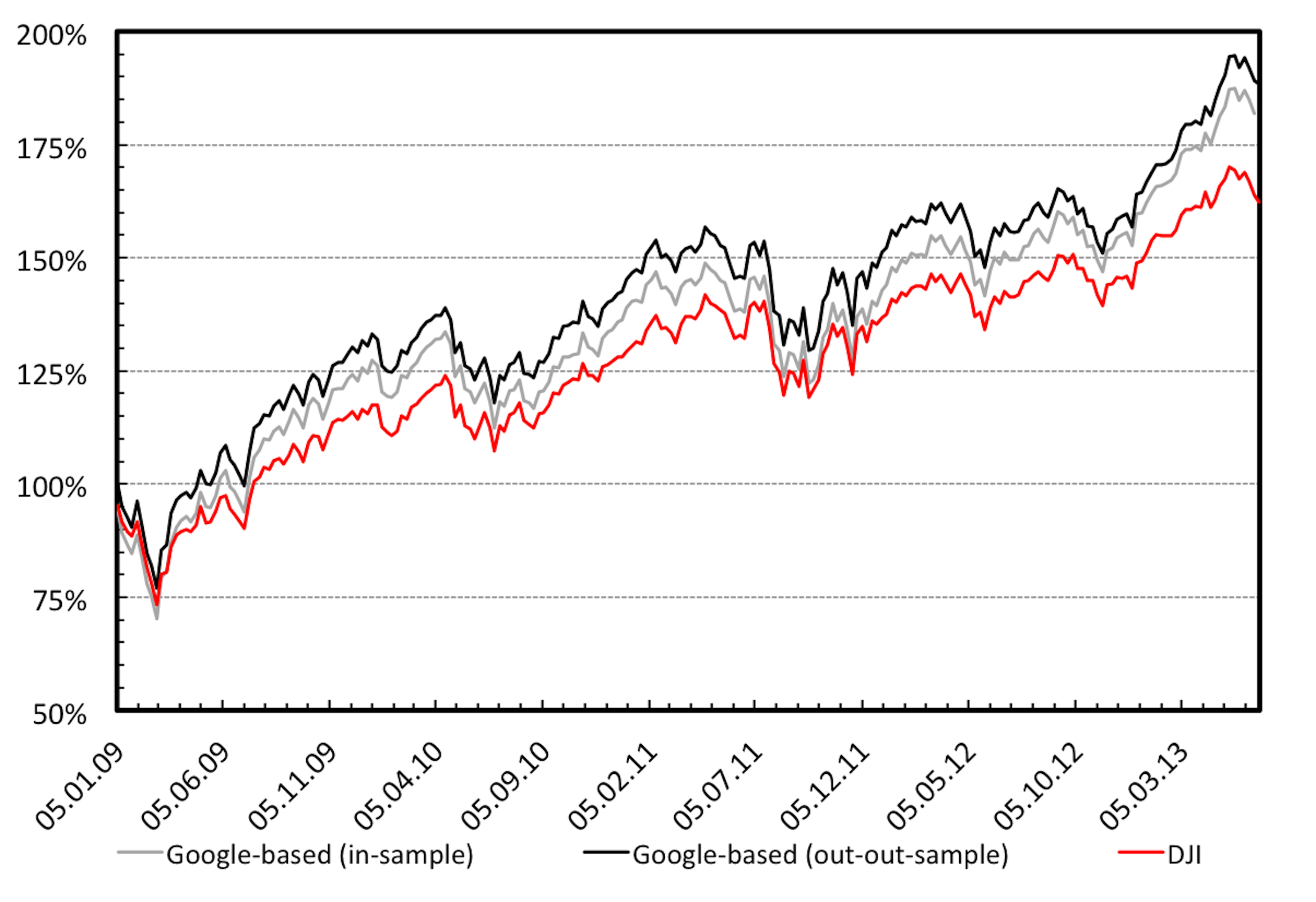}\\
\end{tabular}
\caption{\footnotesize\textbf{Evolution of portfolio value based on \textit{Google Trends} diversification.} Red line represents the evolution of the DJI index, black line shows the performance of the out-of-sample diversification and the grey line illustrates the development of the in-sample diversification approach. Portfolio value is shown on the $y$-axis. On the left panel, the ticker symbol searches are utilized and on the right panel, the combination of the word ``stock'' and the ticker symbol is used. The evolution is shown for the discrimination parameters $\alpha$ which maximize the Sharpe ratio for each scenario. For the practical purposes, a comparison of the black and red lines is essential as it shows how much better off we would be if we applied the \textit{Google Trends} based strategy. For the first case, the search queries based strategy brings the total profit which is more than 4 times higher compared to investing to the DJI index. For the second case, the cumulative profit is more than 0.4 times higher for the search based strategy. \label{fig4}}
\end{figure}

In summary, our results indicate that the search queries from \textit{Google Trends} can be utilized in portfolio selection and risk diversification. Without any information about the correlation structure between components of the DJI index, we were able to construct portfolios which dominate both the uniformly weighted portfolio and the benchmark DJI index. To further illustrate this, we provide Fig. \ref{fig4}. There we show evolution of the Sharpe ratio maximizing portfolios for the in-sample and the out-of-sample strategies compared to the DJI index. It needs to be stressed that the portfolios are not the profit (or average return) maximizing ones but only the Sharpe ratio maximizing ones as the main aim of this report is to discuss the utility of search queries for portfolio optimization and not to find the profit maximizing strategy. The dominance of the profit maximizing portfolios would be even more apparent yet based on different weights.

For the first -- ticker -- strategy, the value of the in-sample portfolio at the end of the analyzed period is 209\% of its initial value which corresponds to a cumulative profit of 109\%. The out-of-sample performance is even better with a 163\% cumulative profit. Compared to the DJI index with a cumulative profit of 38\%, the search-based strategy yields approximately a quadruple profit. Note, however, that neither of the strategies was able to shrink a huge loss linked to the Lehman Brothers bankruptcy in September 2008, which indicates that such an event was practically impossible to diversify out in the U.S. context. Nonetheless, the overall performance of the \textit{Google Trends} based strategy is very promising.

For the second -- stock/ticker -- strategy, a dominance of the search-based portfolios is not so apparent. Even though the cumulative profit of the out-of-sample strategy is equal to 88\% compared to 62\% of the DJI index, the difference is not satisfying mainly due to an issue of transaction costs. The search-based strategies are based on weekly rebalancing of weights in the portfolio which can bring substantial (transaction) costs. An extra profit to cover such costs is needed to justify the more active strategy based on search queries. The difference of 26\% over 4.5 years is not satisfying for that matter. Returning to the previous case based solely on ticker symbols, the difference of 125\% over 8.5 years is much more promising and satisfying. The difference between ticker and ticker/stock strategies is also apparent by comparing their performance after year 2009. The ticker strategy clearly diverges from the values of the DJI index whereas the value of the stock/ticker portfolio remains very close to the index. Overall, the sole ticker symbols seem to bring more reliable information about interest in trading of a specific stock. Most probably, the investors are looking simply for the ticker symbol rather than combining it with the word ``stock'' making the information content in the more complicated query less useful.

We have thus shown that the search queries on \textit{Google Trends} can be successfully utilized for portfolio diversification, which contributes to the contemporary literature mainly dealing with the profit opportunities associated with the search engines and financial markets \cite{Preis2010,Preis2013,Moat2013}.

\section*{Methods}
\subsection*{Data}
We utilize search query series from \textit{Google Trends} website which have several crucial characteristics. Firstly, the series are reported with a weekly (Sunday-Saturday) frequency. For the stock price and returns series, we thus reconstruct the weekly returns to match this pattern. The query series are updated daily for the running week so that there is no lagging issue for the portfolio construction. Connected with the fact that stock markets are closed over weekends, if an investor wants to rebalance his or her portfolio for the next week, he or she has the needed information on Sunday so that the portfolio can be rebalanced efficiently. Secondly, the search volumes are normalized so that each series has a maximum of 100 and the rest of the series is rescaled. However, if more searched terms are looked up on the website, all the series are normalized according to the maximum of all series. Unfortunately, it is only possible to search up to 5 items. For the DJI index with possibly 30 components, we opted for rescaling according to the searched volumes of various groups of five. This way, we are prone to make rounding errors but we assume that these errors are random. Moreover, the \textit{Google Trends} are already rounded themselves so that the issue of rounding cannot be avoided in any way for these series.

As for the DJI index, the components have changed during the analyzed period slightly. Specifically, five companies have been switched since 2005 (see http://www/djiindexes.com for the detailed history). For our analysis, we utilize the structure of the DJI index as of 24.6.2013. As we are not primarily interested in the performance of our portfolio compared to DJI but rather in the potential of search queries in risk diversification, we do not adjust the utilized dataset. Comparison to the DJI index is added just for illustration.


\bibliographystyle{naturemag}

\section*{Acknowledgements}
The support from the Grant Agency of the Czech Republic (GACR) under projects P402/11/0948 and 402/09/0965, Grant Agency of Charles University (GAUK) under project $1110213$ and project SVV 267 504 are gratefully acknowledged.

\section*{Author contributions}
L.K. solely wrote the main manuscript text, prepared the figures and reviewed the manuscript.

\section*{Additional information}
\textbf{Competing financial interests:} The author declares no competing financial interests.\\
\textbf{Data retrieval:} Search volume data were retrieved by accessing the Google Trends website (http://www.google.com/trends) on 26-27 June 2013. DJI stocks component prices were retrieved from http://finance.yahoo.com on 26 June 2013.

\newpage

\end{document}